# Universal Model of Ontogenetic Growth: Substantiation and Development of Schmalhausen's Model

Leonid M. Martyushev, Pavel S. Terentiev

Institute of Industrial Ecology, 20 S. Kovalevskoy Str., Ekaterinburg, 620219, Russia
Ural Federal University, 19 Mira Str., Ekaterinburg, 620002, Russia
leonidmartyushev@gmail.com

## Abstract

A hypothesis of singleness of the growth equation for biological objects on different organizational levels and dimensional analysis are used in order to substantiate Schmalhausen's model of ontogenetic growth (the mass of a growing organism is a power function of time). It is stated that such a model is valid only in the initial period of growth. For the whole period of growth, a generalization of Schmalhausen's model is advanced; it provides the same accuracy as previously known models of quantitative description of kinetic curves. Within the scope of the developed model, a number of interesting results related to an allometry and biological time are obtained.

## 1. Introduction

The study of growth regularities is critical for a number of branches of biology and medicine connected with the investigation of embryonal and post-embryonal development, tissue regeneration, carcinogenesis, aging, etc. Furthermore, these investigations are crucial from the practical point of view primarily in agriculture. The investigation of biological growth is conducted on different levels (cells, tissues, organs, and organism in general) and has many unsolved problems. The search for equations describing the behavior of mass as a function of time during the growth is one of these problems. The literature on this topic is quite extensive and contains multiple growth models [1-3]. The models by Verhulst (the so-called Logistic Model), Gompertz, and Bertalanffy are the most well known. Despite different mathematical form, these models are able to provide rather good quantitative description of the available empirical data that, unfortunately, often are not very accurate. If growth equations are considered only as a suitable tool for describing and predicting the behavior of mass depending on age with a preset accuracy for some biological species, then there is obviously no unambiguous solution for the model-selection problem. This problem is solved using the methods of mathematical statistics and is of applied interest only. However, if a growth equation is considered as a mathematical projection of some universal regularity of biological development, then such a problem becomes significant and interesting from the standpoint of theoretical biology. An assumption of some universal law of development existing in the animated nature leads us to the conclusion of singleness of the growth equation for different biological objects. Herein, we shall assume this hypothesis.

The simplest, but not always indisputable, arguments of biological or chemical origin are most often used as a basis for building models of growth. For example, in the case of the Bertalanffy equation (or modifications thereof), it is considered that the variation of an organism's mass results from two independent processes: synthesis (anabolism) and destruction (catabolism) of the building materials of the body [4]. Each of these processes (summands) is assumed proportional to the organism mass with some different powers. There is a long and controversial discussion regarding the numerical values of these powers, especially in relation to the summand responsible for matter synthesis (2/3 or 3/4) (see, for example, [5,6]). Equations similar to Bertalanffy's equation attempt to give a more rigorous substantiation by using more thorough analysis of energy fluxes that exist in the growth process. Nevertheless, recent papers [7-13] have demonstrated that these substantiations also have numerous complications. The use of energy arguments is certainly a step forward as compared to intuitive assumptions that were predominantly used for building the first growth models. However, consideration of the balance ratios of energy and matter (i.e., corollaries of the first law of thermodynamics) alone without involving the concepts of irreversibility characterized by entropy and growth time (i.e., corollaries of the second law of thermodynamics) is essentially incomplete for such evidently non-equilibrium and dissipative objects as living organisms.

The degree of detail is an important aspect for formulating a growth model. This can be a two-parameter model with one variable of Bertalanffy type [4] or a model with more than a dozen of variables and parameters, for example [7]. The first-type models enable to obtain explicit solutions, qualitatively or even quantitatively describe the behavior of organism mass as a function of time, generally understand the role of these or those factors. The second-type models are very informative but have no explicit solution and focus on numerical calculation. In principle, the models of the second type can provide good quantitative predictions, which fact, in certain cases, allows replacing real experiments or field observations with computer-based virtual experiments. Depending on the problems at hand, models of both the first and the second type are required. It is obvious that the simplest models of the first type are more convenient and useful for identifying the basic regularities of the biological development, because they originally include only the essentials. It is these universal models that we shall consider herein. The type of a mass-time dependence in the initial period of growth is one of the fundamental problems that were not solved in this regard. A number of models provide an exponential dependence, while the others provide a power-law dependence (see, for example, [2]). From the mathematical standpoint, these variants of behavior in time are fundamentally different, which can result in different corollaries relevant for theoretical biology.

So, the problem of building the simplest universal models of ontogenetic growth and analyzing their basic properties is still topical and this issue shall represent the subject hereof.

## 2. Model and its substantiation

In order to build a growth model, we have to rely only on the general laws and methods. A biological system developing in time is extremely complex for description (e.g., complications occur even for the simplest energy calculations [8,9]). The method of dimensional theory is a traditional and reliable method for considering complex systems that are difficult to formalize in mathematical terms [14-16]. This approach allows establishing correct fundamental relations between the quantities determining the process. Conclusions of the dimensional method are sometimes too general and are of no practical importance. However, in certain cases, this method enables to obtain a result practically agreeing with more rigorous and complex approaches. We shall use this method herein.

The change of mass $m$ relative to time $t$ is the traditional unknown quantity in the biological growth equation: $\dot{m}$ (≡d$m$/d$t$). As is demonstrated by multiple experiments and the overwhelming majority of existing theories, this quantity (growth rate) should depend primarily on the mass $m$ of a growing body (organ, etc.). In addition, there should be other quantities determining $\dot{m}$. Let us designate the set of such quantities as $\{a_i\}$. Then, let us take into account the fact that we have assumed the hypothesis of universality meaning that the growth equations for different living beings (animals, mushrooms, plants, etc.) as well as for biological objects on other organizational levels (organs, tissues, etc.) should be similar. So, this universal equation should not contain dimensional constants and variables specific for this or that growing object. For example, for plants, important dimensional parameters and constants could be wavelength of incident radiation and Planck's constant; for bacteria, medium-diffusion coefficients and gravitational acceleration; and for cells and organelles, rate constants of different chemical reactions, etc. Dimensional quantities that are of fundamental importance for the growth of some systems can be only secondary or meaningless for the growth of other systems. Therefore, the equation universality excludes from consideration all the quantities that could influence growth rate in each particular case. At the same time, based on the dimensionality considerations, $\{a_i\}$ *should* include a quantity with time dimensionality. Age seems as the primary choice for such a quantity (let us designate it in the same way as time[1] $t$). Age, i.e. the time period from birth (origin) to some moment of development, can be introduced for any multicellular biological object (as opposed to, for instance, molecules and atoms); and the influence of age on $\dot{m}$ is obvious[2]. Consequently, age undoubtedly fits to be the required dimensional variable for the universal growth equation. Factors describing the change in properties of a medium where the growth occurs could serve as other quantities related to time. Let us first consider the case when a medium where the growth occurs does not change its properties with time. Obviously, such an approach is true for the initial period of any growth (embryonal, post-embryonal). As a result, we come to the conclusion that, for the case at hand, $\dot{m}$ is a function of $m$ and $t$ only, which, according to the dimensionality theory, has the form:

$$\dot{m}=a\cdot m/t, \tag{2.1}$$

where $a$ is some dimensionless constant. Equation (2.1) is easily transformed to the following:

$$m=C\cdot t^{a}, \tag{2.2}$$

where $C$ is some integration constant.

[1] Herein, this will not lead to confusion.

We have obtained growth equations (2.1) and (2.2) based on two assumptions: 1) of the universality of development processes for various biological objects, 2) that changes in a medium can be neglected. It should be noted that exactly these equations were empirically obtained on the basis of experiments (first, using the growth data of chicken embryos, then using a number of other biological species both for embryonal and post-embryonal growth) by I. Schmalhausen in 1925-1935 [18-21]. According to Schmalhausen, the power law (2.2) provides considerably better results than the exponential one, especially for embryonal growth. This growth model was also confirmed in the experiments of other researchers (see, for example, [22, 23]). The value of the coefficient $a$ found in these experiments ranged from 0.1 to 10. Schmalhausen connected his power-law empirical model with the role of differentiated cells and with their number changing with age. According to Schmalhausen, the exponential growth can be observed only for the growth of organisms without differentiation (for instance, bacteria), whereas the power-law growth (2.2) occurs for the growth of organisms with differentiation (for example, vertebrates). The arguments of Schmalhausen, which were mainly of qualitative nature and based on a large number of assumptions, did not attract attention of biologists both during his lifetime and afterwards. As far as we know, formulae (2.1) and (2.2) were not discussed from the standpoint of theoretical substantiation for several decades. However, a paper [24] has recently appeared considering a number of questions related to the physiological and physical time of living systems. The authors of this study use an interesting but unobvious hypothesis that fluctuations in the total body mass are described by the scaling probability density and obtain a number of formulae that, as is noted in our commentary [25], lead directly to (2.1) and (2.2.).

---

[2] Mutations and waste products accumulate in an organism (organ) with time, which undoubtedly influences the mass increment. In particular, it is well known that the wound healing rate considerably changes with age (e.g., according to the data [17], the relative rate of wound healing decreases with age approximately following a hyperbolic law). We can make a lot of other examples showing the influence of age on the mass changing rate.

Schmalhausen's empirically obtained formulae (2.1) and (2.2) were criticized. There were two main objections. The first objection (see, e.g., [26]) is that the presence of an explicit time variable in (2.1), is unnecessary as growth rate should be determined by mass only. As a reply, in addition to the arguments that we used in the deduction of (2.1), it can be noted that if the time is expressed through (2.2) and introduced into (2.1), then:

$$\dot{m} = a \cdot C^{1/a} \cdot m^{(a-1)/a} \qquad (2.3)$$

Thus, the time, if necessary, can always be excluded from (2.1).

The second objection is connected with the description of animal-growth data using Eq. (2.2). So, it is obvious that complex curves – for instance, the so-called sigmoids (*S*-shaped) – cannot be described using the dependence of (2.2) type. Schmalhausen considered that it is incorrect to describe a growth curve over the whole time interval (in this regard, he concurred with, for example, Brody [26]) and the growth of a living organism needs to be divided into stages to which (2.2) shall be subsequently applied. According to Schmalhausen, these stages have to be natural and determined by the biological development specifics of this or that species. For example, the embryonal growth period is a special stage; the period from birth to lactation end of a mammal can be another stage, etc. A break is often observed in the growth curves, which can serve as a signal of the beginning of a new stage. However, there is not always a sharp break in the growth curves. In this case, the growth stages are difficult to define and different researchers (depending on the used mathematical and/or biological arguments and preferences) may select different borders of the stages or even different number thereof. Thus, for example, an *S*-shaped curve can be divided into two stages – before inflection (accelerating growth) and after inflection (decelerating growth) – or into three stages, etc. The mentioned circumstance was a significant restriction for the widespread use of formulae (2.1) and (2.2) and caused criticism of Schmalhausen's approach. After obtaining (2.1) from the studies of embryonal growth, Schmalhausen considered that it is both applicable to other stages (e.g., the post-embryonal one) and completely describes the ***whole*** period of growth for any stage. As follows from our deduction of (2.1) using the dimensionality theory, the latter is incorrect: Eq.(2.1) describes only the initial period of growth when conditions (an environment) around a growing organism can be considered invariable. An abrupt change in the conditions accompanied by a break of the growth curve, indeed, separates different stages of growth, and formulae of (2.1) or (2.2) type need to be applied to each stage (to their initial periods). However, even during one stage of growth, a gradual change in the conditions influencing the growth is possible; the more time passes from the initial moment of the stage under consideration, the stronger the influence should be. As a result, formulae (2.1) and (2.2) become poorly applicable.

Based on the above, these formulae should be generalized for the case when growth conditions change (for example, because of the properties of an environment around a growing body). It is the most obvious and the simplest way to assume that the change occurs only due to the growth itself and is proportional only to the mass of the growing body. As a result, (2.1) can be rewritten in the form:

$$\dot{m} = a \cdot m/t - b \cdot m \ , \qquad (2.4)$$

where $b$ is some positive constant with reverse-time dimensionality. By integrating (2.4), we can obtain:

$$m = C \cdot t^{a} \cdot exp(-b \cdot t) \qquad (2.5)$$

Formulae (2.4) and (2.5) are the development of Schmalhausen's model[3] and describe (depending on the ratio of the coefficients $a$ and $b$) both $S$-shaped growth dependencies as a whole and accelerating or decelerating growth only. As is seen from (2.5), the acceleration (for $a>0$) occurs in line with the power-law dependence, whereas the decrease of the growth rate occurs in line with the exponential dependence. It should be additionally emphasized that while we consider the power-law behavior of mass with time during the growth as rather a universal phenomenon in nature, the exponential decrease is considered as a common but not comprehensive phenomenon. As a consequence, $m(t)$ is a skewed S-shaped curve. This is an advantage of the proposed model (as compared, for instance, with the Verhulst model) because, as is known from the experimental data, the period of growth-rate decrease is usually longer than the period of its increase [27].

Let us compare (2.4) with Bertalanffy's model. As is known, $\dot{m} = \alpha \cdot m^{\gamma} - \beta \cdot m$ in Bertalanffy's model, where $\alpha$, $\beta$, $\gamma$ are some coefficients (with $\gamma$ in the range of 2/3 to 3/4, according to the measurements and theoretical reasoning [5]). Since $\gamma \approx 1$, the explicit dependence of the first summand of DS model on time is the only difference between the two growth models. Due to this dependence, the specific growth rate in (2.4) is always a decreasing power-law function of time[4]. Such an explicit dependence leads to an important property of DS model, which is missing in Bertalanffy's model (and many other widely used models). So, according to (2.4), the requirement of $\dot{m}=0$ allows finding the time of growth cessation $t^*$ equal to $a/b$. This time has a finite defined value (in the models of Bertalanffy type, the stop of growth occurs only asymptotically at an infinite time). According to (2.5), the body mass at the moment of growth stop $t^*$ equals to $m^* = C \cdot (a/b)^{a} \cdot exp(-a)$. By using the introduced scales $t^*$ and $m^*$, let us nondimensionalize (2.4) and (2.5) and rewrite them in the dimensionless variables of mass $\tilde{m}$ and time $\tilde{t}$ :

[3] Therefore, let us shortly refer to a new model as DS model.

[4] Thus, one of the main laws (according to Ref.[28]) of biological growth is immediately satisfied: "the specific growth-rate declines more and more slowly as the organism increases in age".

$$\dot{\tilde{m}} = a \cdot \tilde{m} \cdot (1-\tilde{t})/\tilde{t} \quad (2.6)$$

$$\tilde{m} = \tilde{t}^{a} \cdot exp(a \cdot (1-\tilde{t})) \quad (2.7)$$

As is seen from (2.6) and (2.7), after nondimensionalization, the relationship between mass and time over the *whole* time period of growth depends only on the coefficient *a*, which, according to the discussion above, is directly connected with the initial period of biological growth. Formulae (2.6) and (2.7) can be convenient for describing the growth data. The quantities $t^*$ and $m^*$ can be both determined immediately from the experimental data (for example, in case a growth stop is observed in the experiment) and calculated using the found values of *a* and *b* (for example, in case a growth stop is not observed during the whole lifetime as, for instance, for fishes).

## 3. Description of experimental data using DS model

It was proposed above to describe biological growth using DS model (2.4)-(2.7). We put forward this model generalizing Schmalhausen's empirical model in order to describe one natural stage of an organism's growth[5]. Based on the available experimental data with the dependence of living-organism mass on time, let us check the model for its abilities to describe growth in terms of quantity. Data related both to animals and to plants with one explicit growth stage was used (see examples in Fig.3.1). The embryonal growth was not considered because it has sufficiently many evidences of good description obtained using the Schmalhausen-type model; and consequently, description using a model with an additional parameter (Eq.(2.5)) should lead only to better description. In order to select the literature with the data for quantitative analysis, we relied on the following: 1) availability of at least ten measurements of mass describing the whole growth period; and 2) accuracy (for each species, information was obtained based on the study of at least several dozens of individuals). The data used in the study is detailed in Table 3.1. This table gives the values of the coefficients (*C, a,* and *b*) obtained on the basis of approximation (2.5). Table 3.2 contains the coefficients (coefficient of determination and normalized root-mean-square error) corresponding to the quality of fitting the experimental points using the model. All the quantities (Table 3.1 and 3.2) were determined in Curve Fitting Toolbox (MatLab). As is seen from the given information, equation (2.5) describes experimental data (also see Fig. 3.1) as good as the traditional models by Verhulst, Gompertz, and Bertalanffy. Let us note that all the four models describe the used experimental data with almost the same accuracy. Consequently, as was mentioned above, in order to select a model, one should rely on its validity with regard to commonly accepted principles and consistency of its foundations. Fig. 3.2 shows all the used experimental data subject to preliminary nondimensionalization with the scales $t^*$ and $m^*$. The data was plotted in the transformed coordinates, where, according to (2.7), the dependence of dimensionless mass on dimensionless time should have the simplest linear form. As is seen in Fig. 3.2, the points are indeed positioned rather compactly along the line.

[5] According to (2.2) and (2.5), the zero mass corresponds to the initial (zero) time of a growth stage under consideration. Here, mass shall mean only the mass during the growth for the stage at hand (in case there are previous stages, this mass will certainly differ from the actual mass of an organism). The growth time is a sum of full times of each growth stage, and the mass of the grown organism is composed of the values of mass reached on each of the stages.

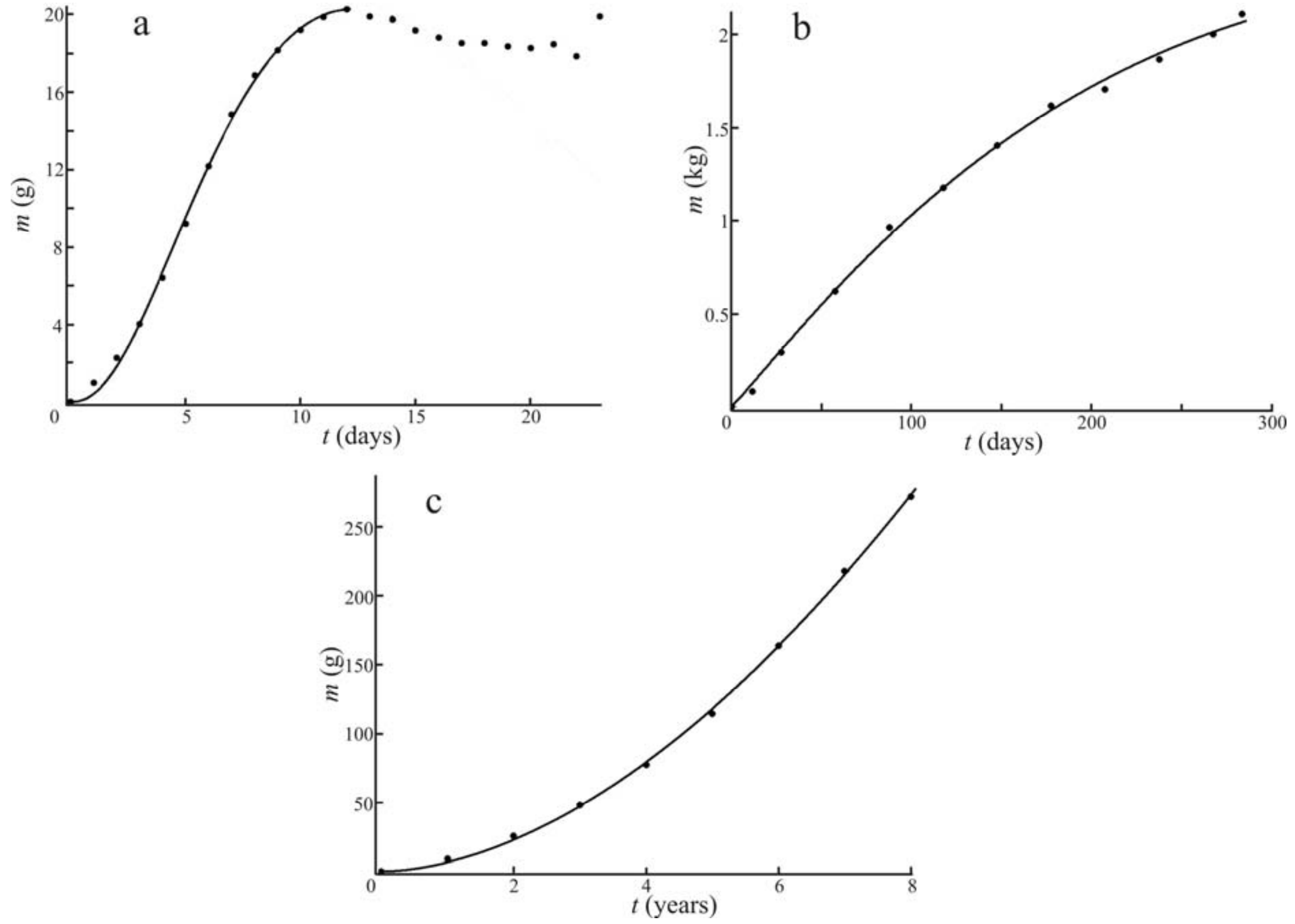

**Fig.3.1.** Examples of the basic types of the experimental dependences of mass on time $m(t)$ considered herein: (a) with the pronounced saturation and the subsequent decrease of $m(t)$ (Tree Swallow [37]); (b) continuous growth of $m(t)$ with the decrease of the mass increment rate (Swamp Rabbit, [42]); (c) continuous growth of $m(t)$ with the increase of the mass increment rate (White Bream, [32]). The points are the results of the experimental measurements, the lines are the results of the approximation using DS model (for the approximation of the data on Tree Swallow, only the points corresponding to the mass-increase stage were used).

**Table 3.1.** The parameters of DS model for some organisms. The experimental dependences of mass on time taken from the papers specified in the Reference column.

| **Name** | ***C*** | ***a*** | ***b*** | **Reference** |
|---|---|---|---|---|
| Yorkshire Fog (*Holcus lanatus*) | $2.5\pm0.5 \cdot 10^{-6}$ [mg/day] | 5.3±0.8 | 0.17±0.03[1/day] | Shipley, B. (1996) [29] |
| Gooseberry (*Actinidia chinensis*) | 7±5[g/week] | 0.9±0.3 | 0.04±0.02[1/week] | Pratt, H.K. (1974) [30] |
| Cockroach (*Periplaneta americana*) | $1\pm1 \cdot 10^{-4}$[mg/year] | 2.75±0.06 | 0 | Gier, H.T. (1947) [31] |
| Common Bream (*Abramis brama*) | 2±1[g/year] | 4.1±0.9 | 0.4±0.1[1/year] | Specziár, A. (1997) [32] |
| White Bream (*Blicca bjoerkna*) | 7±2[g/year] | 1.79±0.07 | 0 | Specziar, A. (1997) [32] |
| Rock Lobster (*Jasus edwardsii*) | 0.25±0.01[kg/year] | 1.11±0.03 | 0.05±0.05[1/year] | McGarvey, R.(1999)[33] |
| Norway Lobster (*Nephrops norvegicus*) | 0.005±0.001[kg/year] | 1.46±0.09 | 0 | Smith, I.P. (2008) [34] |
| Clifford's Snake (*Spalerosophis cliffordi*) | 15±2[g/year] | 1.18±0.06 | 0 | Dmi'el, R. (1967) [35] |
| Painted Turtle (*Chrysemys picta*) | 43±9[g/year] | 1.5±0.4 | 0.20±0.07[1/year] | Wilbur, H.M. (1975)[36] |
| Tree Swallow (*Iridoprocne bicolor*) | 0.5±0.1[g/day] | 2.5±0.3 | 0.20±0.03[1/day] | Zach, R. (1982) [37] |
| Cassin's Auklet (*Ptychoramphus aleuticus*) | 0.8±0.2[g/day] | 1.9±0.1 | 0.043±0.006[1/day] | Thoresen, A.C.(1964) [38] |
| Atlantic Gannet (*Sula bassana*) | 0.7±0.7[g/day] | 2.7±0.3 | 0.039±0.007[1/day] | Nelson, J.B. (1964) [39] |
| Black browed albatross (*Thalassarche melanophris*) | 0.010±0.005[kg/day] | 1.7±0.2 | 0.016±0.003[1/day] | Ricketts, C. (1981) [40] |
| Antarctic Fur Seal (*Arctocephalus gazella*) | 8±8[kg/year] | 2±1 | 0.1±0.2[1/year] | Payne, M.R. (1979) [41] |
| Swamp Rabbit (*Sylvilagus aquaticus*) | 7±4[g/day] | 1.1±0.1 | 0.003±0.001[1/day] | Sorensen, M.F. (1968) [42] |
| Shrew (*Cinereus ohioensis*) | 0.07±0.04[g/day] | 2.1±0.4 | 0.11±0.03[1/day] | Forsyth, D.J. (1976) [43] |
| Soft-fur Rat (*Millardia meltada*) | 6.0±0.8[g/week] | 1.6±0.1 | 0.13±0.02[1/week] | Yosida, T.H. (1978) [44] |
| Goat (*Capra hircus*) | 0.2±0.1[kg/day] | 0.90±0.09 | $6\pm1 \cdot 10^{-4}$[1/day] | Zullinger, E.M. (1984) [45] |
| Cow (*Bos taurus*) | 0.4±0.4[kg/day] | 1.1±0.2 | $6\pm4 \cdot 10^{-4}$[1/day] | Brody, S. (1945) [26] |

**Table 3.2.** The quality of curve fitting of the experimental data (see Tabl. 3.1) using DS model proposed herein as well as the models of Bertalanffy $(a\cdot(1\text{-}\exp(-c\cdot b\cdot t))/b)^{1/c}$, Gompertz $c\cdot\exp(-a\cdot(\exp(-b\cdot t)))$, Verhulst $c/(1+a\cdot\exp(-b\cdot t))$ (the parameters *a,b,c* were fitted).

| Name | R-squard | | | | Normalized root mean square error (x100) | | | |
|---|---|---|---|---|---|---|---|---|
| | **DS-model** | **Bertalanffy** | **Gompertz** | **Verhulst** | **DS-model** | **Bertalanffy** | **Gompertz** | **Verhulst** |
| Yorkshire Fog | 0.9915 | 0.9961 | 0.9962 | 0.9951 | 4 | 3 | 3 | 3 |
| Gooseberry | 0.9491 | 0.9435 | 0.9297 | 0.9310 | 7 | 7 | 8 | 8 |
| Cockroach | 0.9997 | 0.9998 | 0.9997 | 0.9987 | 1 | 1 | 1 | 1 |
| Common Bream | 0.9978 | 0.9972 | 0.9981 | 0.9994 | 2 | 2 | 2 | 1 |
| White Bream | 0.9995 | 0.9995 | 0.9995 | 0.9980 | 1 | 1 | 1 | 2 |
| Rock Lobster | 0.9999 | 0.9998 | 0.9956 | 0.9873 | 1 | 1 | 2 | 4 |
| Norway Lobster | 0.9994 | 0.9981 | 0.9825 | 0.9575 | 2 | 2 | 6 | 9 |
| Clifford's Snake | 0.9998 | 0.9997 | 0.9988 | 0.9942 | 2 | 1 | 1 | 3 |
| Painted Turtle | 0.9957 | 0.9997 | 0.9974 | 0.9894 | 3 | 1 | 2 | 4 |
| Tree Swallow | 0.9987 | 0.9972 | 0.9989 | 0.9987 | 1 | 2 | 1 | 1 |
| Cassin's Auklet | 0.9982 | 0.9979 | 0.9980 | 0.9941 | 1 | 2 | 2 | 3 |
| Atlantic Gannet | 0.9885 | 0.9857 | 0.9869 | 0.9855 | 4 | 4 | 4 | 4 |
| Albatross | 0.9974 | 0.9979 | 0.9970 | 0.9907 | 2 | 2 | 2 | 4 |
| Seal | 0.9714 | 0.9707 | 0.9764 | 0.9806 | 7 | 7 | 7 | 6 |
| Swamp Rabbit | 0.9988 | 0.9991 | 0.9928 | 0.9818 | 1 | 1 | 3 | 5 |
| Shrew | 0.9974 | 0.9937 | 0.9956 | 0.9951 | 2 | 4 | 3 | 3 |
| Soft-fur Rat | 0.9993 | 0.9991 | 0.9984 | 0.9921 | 1 | 1 | 2 | 4 |
| Goat | 0.9979 | 0.9971 | 0.9881 | 0.9751 | 2 | 2 | 4 | 6 |
| Cow | 0.9893 | 0.9961 | 0.9884 | 0.9755 | 4 | 2 | 4 | 6 |

As follows from the previous section, the dimensionless coefficient *a* is the most significant for the growth model under consideration. Its frequency bar chart is shown in the insert to Fig. 3.2. According to the processed data, the most probable values of *a* are approximately within the range of 1 to 2. A relatively small number of considered species and the accuracy of data prevent us from making a conclusion of the dependence or independence of this coefficient from these or those characteristics.

According to the given data (Table 3.1), the coefficient *b* proves to differ from zero in 14 cases out of 19 that we considered. Furthermore, in one case (Antarctic Fur Seal), it cannot be definitely stated based on the available data that the coefficient has the zero value. Thus, according to the calculations, the inclusion of the coefficient *b* in DS model is justified.

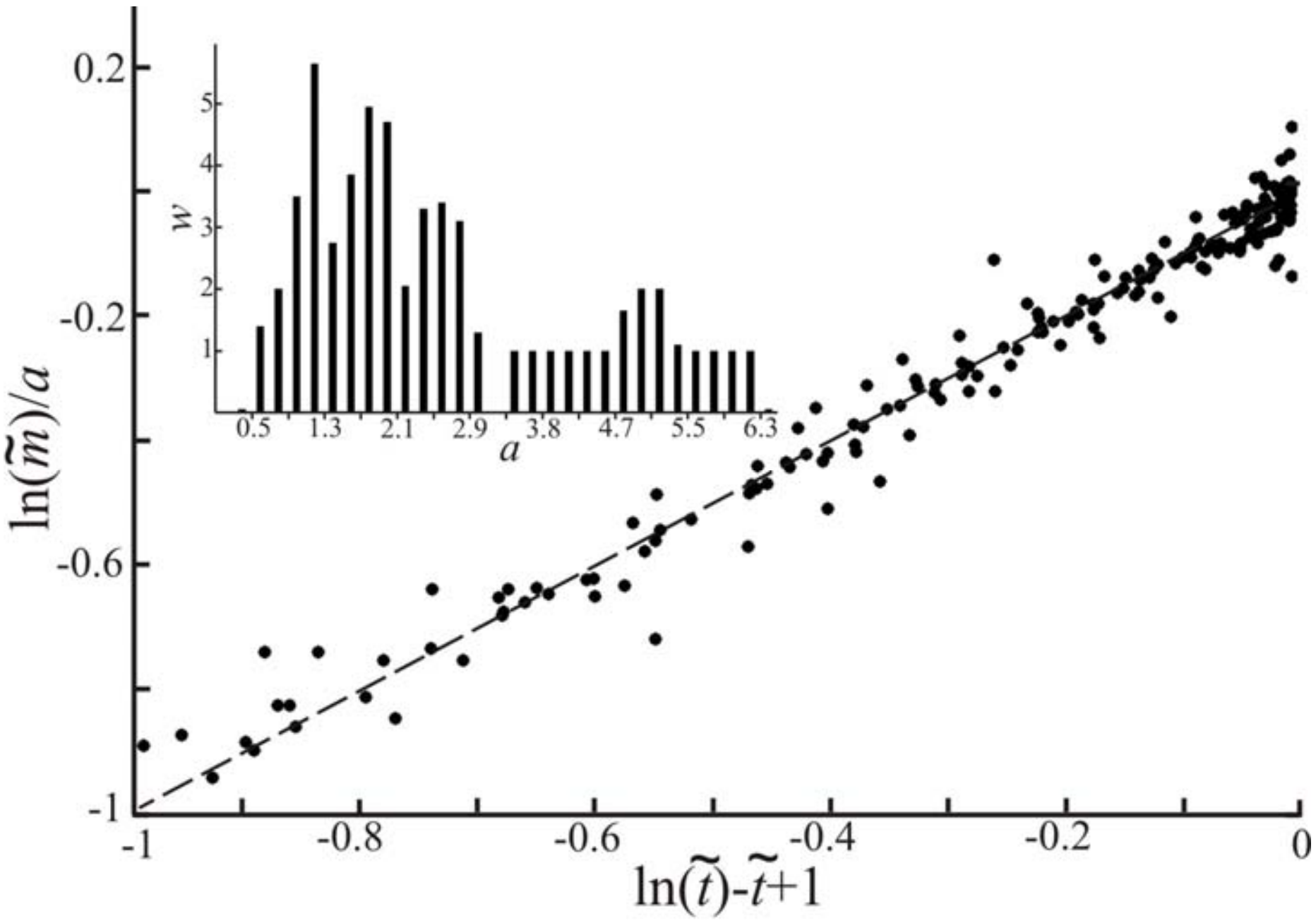

Fig.3.2. Dependence of the dimensionless mass $\tilde{m}$ on the dimensionless time $\tilde{t}$ plotted on the transformed coordinates. The data of all 19 investigated species is presented. The coordinates are transformed such that the dependence $\tilde{m}(\tilde{t})$ determined by Eq. (2.7) has the simplest form: it is shown with the dashed line in the figure. The masses and times were preliminary non-dimensionalized by the whole growth time $t^*$ and the maximum mass $m^*$ (see the text for details). The individual parameters *a* and *b* found for the organisms at hand and given in Tabl. 3.1 were used for non-dimensionalization and the transformation of axes. The frequency bar chart *w* of the *a* values distribution (Tabl. 3.1) is shown in the insert to the figure.

## 4. Relation of the developed model to other results

In this section, we shall list and briefly discuss a number of interesting relations between the proposed model (2.1) – (2.5) and the results of previously published papers.

1. According to formulae (2.1) and (2.2), the mass of a growing organism is a homogeneous function of the degree *a.* Consequently, the dependence of mass on time (2.2) is scale-invariant (self-similar), i.e. its form does not change when observed in different time ranges (e.g., periods from birth to age of 1 sec. or up to 100 sec.). This property is a corollary of the absence of the characteristic time scale resulting from the assumed universality of dependence (2.1). The homogeneity of the function $m(t)$ leads to the following important property. Indeed, let $m_1(t)$ and $m_2(t)$ be the masses of two simultaneously growing parts of some organism (we shall designate the parameters related to these organisms with the indices 1 and 2). Then, according to (2.1):

$$\dot{m}_1/(a_1 m_1) = \dot{m}_2/(a_2 m_2) \quad \text{or}$$

$$\dot{m}_1/m_1 = (a_1/a_2)\cdot \dot{m}_2/m_2 \tag{3.1}$$

This law (law of ontogenetic allometry) of proportionality of relative changes in the masses of growing parts of an organism was empirically proposed by J. Huxley (1932) [46][6]. The mentioned simplest relation between the power-law dependence *m*(*t*) and the allometric law was stated as early as in 1930 by I. Schmalhausen [18-21]. It should be noted that other common dependencies of mass on time (for example, those by Verhulst and Bertalanffy) lead to the allometric law of (3.1) type only in the case of significant additional assumptions regarding the values of the coefficients included in these growth equations. There are no restrictions imposed on the values of $a_1$ and $a_2$ in the given conclusion. As was mentioned by Huxley [46] and confirmed in subsequent empirical studies [47-50], the relations of (3.1) type are often valid not for the whole growth stage but only for some of its periods (growth of the brain relative to the body is one of the well-known examples). In this regard, let us remind that, as opposed to Schmalhausen, we consider equations (2.1) and (2.2) as valid only for the initial period of some natural stage of growth. For an arbitrary moment of time of the selected growth stage, equations (2.4) and (2.5) are proposed. By using (2.4), let us again consider simultaneous growth of two body parts. In this case, after simple transformations, the following can be obtained:

$$\dot{m}_1 / m_1 = (a_1 / a_2) \cdot \dot{m}_2 / m_2 + \lambda \qquad (3.2)$$

or

$$m_1 = \widetilde{C} \cdot m_2^{(a_1/a_2)} \exp(\lambda t), \qquad (3.3)$$

where $\widetilde{C}$ is some constant and $\lambda=(a_1b_2-a_2b_1)/a_2$ or, using the above introduced time scale:

$$\lambda = a_1(t_1^* - t_2^*)/(t_1^* \cdot t_2^*) \qquad (3.4)$$

Thus, the formula of simple allometry (3.1) according to the model considered herein is valid either if the growth is described with the power-law function *m*(*t*) (i.e., if $b=0$) or if growth times are the same for different body parts ($t_1^* = t_2^*$). In other cases, according to the developed model, the generalized law of allometry of (3.2) or (3.3) is true. The laws of allometry are traditionally used not only to describe the development of an organism's parts in the course of its individual development (this is the so-called ontogenetic allometry) [47-50]. In our opinion, the application of (3.2)-(3.4) for studying the behavior of mass during the development of either individuals of one species on the same developmental stage (static allometry) or individuals of different species (evolutionary allometry) may be the most interesting and useful.

2. As is known, the astronomical time [7] *t* often proves to be an inconvenient quantity for analyzing growth regularities in biology (see, for example, [17, 24, 51-57]). This is because the duration of any development stage can be considerably different[8] in terms of $t$ even for genetically close species [51, 52]. Therefore, it is very complicated to identify some laws of development. A convenient and universal metric of time related to internal processes in an organism under study can allow finding some common features and laws of growth that would be indistinct in the case of the traditional astronomical metric. The search for such a metric has been conducted for a long time [17, 51-57]. This quantity was called the developmental[9] time $\tau$. There were several such quantities proposed in the literature. So, in one of the first papers [17], a unit of biological time was represented by the time of wound healing, whereas in the large experimental investigations [51], it was the time between the first and second cleavage divisions. It is unlikely that the mentioned metrics can be convenient and universal for studying growth on various organizational levels. In this regard, the quantities related to metabolism and energy dissipation may be more useful [24, 53-57]. J. Reiss uses the quantity related to mass-specific metabolic rate as a temporary metric in his paper [57]. As is known, entropy production density is immediately related to this quantity [58-61] (so, for an organism that has stopped growing and is at rest, the only difference between them is the temperature multiplier). From the standpoint of modern nonequilibrium thermodynamics, entropy production is the most fundamental quantity characterizing irreversible changes. Historically, this value is used in physics in order to characterize the direction of time. Therefore, it is more logical to select it as a time metric than the specific metabolic rate. Such conclusions are made, in particular, by the authors of [56]. However, no mathematical development of this idea is contained in their paper. Let us show what this assumption may lead to based on the above growth model.

---

[6] The existence of allometry relationship was also discussed before; for example, G. Cuvier introduced the principle of correlation of an organism's parts (1798). However, these principles were speculative and not expressed quantitatively (mathematically).

[7] There are other synonyms – chronological, extrinsic, physical, and clock time – used in the papers.

Let us introduce a change (differential) of the developmental time $\mathrm{d}\tau$ through a change of the astronomical time $\mathrm{d}t$ as:

$$\mathrm{d}\tau \propto \mathrm{d}S_{\mathrm{diss}}=\sigma(t)\mathrm{d}t, \tag{3.5}$$

where $\mathrm{d}S_{\mathrm{diss}}$ is the irreversible (dissipative) change of the body entropy, $\sigma(t)$ is the body-entropy production density (or, in other words, the relation between specific[10] heat production and temperature of a body). A number of studies are dedicated to the measurement/calculation of the entropy production density of living beings. There are both rather laborious quantitative methods connected with the immediate measurement/calculation of the heat generated from surfaces with known temperatures and simpler and qualitative methods based, for example, on the calculation of the oxygen consumed by a body [58-62]. As is well known, entropy production is a strictly positive quantity and, consequently, the introduced developmental time is also positive.

---

[8] The temperature of a body and of the environment where the growth occurs, as well as the mass of the growing body, etc. have a significant impact.

[9] There are other synonyms in the literature, such as physiological, intrinsic, and biological.

[10] Per unit volume or mass (the latter is more suitable for the problems of biology).

There is plenty of information regarding the properties of entropy production in nonequilibrium processes (the growth considered herein is their special case). In particular, a system is initially developed such that the relationship between the cause and the response of this nonequilibrium system is established in order to maximize the entropy production (see, for example, [63, 64]). Further, after establishing basic relations between the cause and the response (i.e., establishing the functional dependence between the thermodynamic forces and fluxes, if the terms of nonequilibrium thermodynamics are used), processes leading to the optimization of dissipative losses, including the minimization of the entropy production, may occur in the growing system [59-61, 63, 64].

The entropy production can be calculated for various processes related to vital functions and development of an organism. On the basis of the subject hereof, let us consider entropy production as connected only with the growth of an organism; and consequently, we shall determine the biological time (its metric) relative to the growing (increasing mass with age) organism.

It is known (see, for example, [61]) that the entropy production density related to the growth is proportional to the specific change of the body mass[11], i.e.:

$$\sigma_{growth} \propto \dot{m}/m \qquad (3.6)$$

As a result, it follows from (3.5)-(3.6) that:

$$d\tau = \Theta\ (\dot{m}/m)\ dt = \Theta\ dm/m, \qquad (3.7)$$

where $\Theta$ is some quantity depending on the body temperature in the general case. By neglecting the change of the body temperature during the growth, we shall obtain:

$$\tau = \Theta\ ln(m/m_0) + \Theta_1, \qquad (3.8)$$

where $\Theta_1$ is the constant resulting from the integration of (3.7) and $m_0$ is the constant with the mass dimensionality introduced for convenience and nondimensionalization of the quantity under the logarithm. If we consider that $m_0$ is the mass of a growing body at the initial moment of time ($t_0$) of a natural stage of growth under study, then $\Theta_1=0$. By taking into account the fact that Eq. (2.2) is valid for the initial period of growth, we shall obtain:

$$\tau = a \cdot \Theta\ ln(t/t_0) + \Theta_1, \qquad (3.9)$$

The relation between the developmental and astronomical times (3.9) obtained above is very interesting. Indeed, with the increase of an organism's age, every following unit of the astronomical time corresponds to an ever-decreasing value of the biological time. In other words, the use of the growing organism's intrinsic (biological) clock leads to the seeming acceleration of the astronomical (physical) time flow. So, the older the organism, the faster the acceleration (an interval of the intrinsic time perceived by the organism contains more amount of the physical time). G. Backman postulated a formula similar to (3.9) connecting two times on the basis of numerous investigations of the animal and plant growth in 1943 [27]. It should be additionally noted that the obtained law is reminiscent of the Weber–Fechner law stating that subjective sensation is proportional to the logarithm of stimulus intensity. In our case, the intrinsic time (to a certain extent it can be even called the subjective mental time) is a logarithm of the extrinsic time (astronomical time). Such an extrapolation of the Weber–Fechner law, where physical stimulus means a temporary “impact" rather than only light, sound, etc., can be found in the paper [65] (see also references therein). At the same time, the obtained dependence (3.8) and (3.9) is not commonly accepted: for example, these equations contradict some results of [24] where physiological time is considered to be a homogeneous function of mass.

[11] Evidently, such a formula can by used only in case $\dot{m} \geq 0$.

The last consideration shows a number of restrictions of the law (3.9) that were not previously noted in the literature. Any deviation from the above assumptions will result in considerable complication of (3.9). So, consideration of the growth in an arbitrary time period rather than the initial one can be the simplest complication. In this case, according to (2.5) and (3.8), we shall have:

$$\tau = a \cdot \Theta \cdot ln(t/t_0) - \Theta \cdot b \cdot (t - t_0) + \Theta_1 \qquad (3.10)$$

In connection with the obtained formula and the addition of the second negative summand therein, let us again emphasize that, in accordance with the metric introduced in (3.5), internal time can only be a strictly positive quantity. Therefore, calculations using approximated formulae like (3.10) have to be performed very carefully. If negative values of $\tau$ are obtained, this will only mean that the formula (3.10) is outside its applicability domain.

3. DS model (Eq. (2.5)) proposed herein for describing the dependence of mass on time for living systems was previously successfully used for describing nonequilibrium crystallization of ammonium chloride from an aqueous solution in a thin (quasi-two-dimensional) capillary [66, 67]. In these experiments, dendrite and other intricate shapes of growth occurred during the crystallization. It proved to be that Eq. (2.5) was equally good for describing the growth of both individual dendrite branches and the dendrite in general. The fact that the function of (2.5) type was good at describing *S*-shaped kinetic curves during the crystallization is apparently not so surprising. There is the other fact that matters. According to the results of the measurements, the values of the coefficient *a* for various dendrites, their branches (as well for other nonequilibrium shapes of growth) were in the range of 1.5 to 2 [66-67]. So, the values of the coefficient *a* prove to be close for rather dissimilar growing systems, both living (see Table 3.1 and Figure 3.2) and inorganic. The found coincidence requires further study.

## 5. Conclusion

Herein, a power-law dependence of an organism's mass on time was obtained using the dimensional analysis and the hypothesis of growth universality on different scales. Based on this dependence, which is valid only in the beginning of the growth, the critical issues (allometry and biological time) that are traditionally related to biological development were discussed. It is shown that the simplest generalization of the obtained power-law dependence can quantitatively describe the whole natural stage of growth and provide the same accuracy as traditionally used models of ontogenetic growth.

The present paper gives rise to a number of interesting questions/problems requiring solution in the future. Let us enumerate two of them. 1) Is the power exponent of time $a$ universal? In other words, whether it has different or identical values for different parts of a growing organism, for animals belonging to one or different species. A large amount of data on the dependence of an organism's mass on time needs to be thoroughly processed using modern statistic methods in order to solve this problem. Further, special requirements to the quality of source data have to be set in terms of its credibility and accuracy. The investigation conducted herein and based on 19 representatives of the living world enabled us to define the range of values of this coefficient only approximately. 2) The introduction of biological time using entropy production as proposed herein offers interesting challenges for research. The point is that the thermodynamic entropy production is currently a sufficiently well studied quantity. As a consequence, the extension of a number of properties of this thermodynamic quantity (through biological time) to the organism development can prove to be very efficient, in particular, as related to the criterion of coexistence of different organisms (organs) during the growth, to the optimality and efficiency of the development, and to the direction and rate of biological evolution. There are a number of studies conducted in this area (see, for example, [24, 60, 63, 64, 68, 69]). However, we believe that this subject is currently almost uninvestigated despite being very deep and interesting.